\documentclass[12pt,aps,floats]{revtex4}

\usepackage{epsfig}

\def\simgt{\stackrel{>}{{}_\sim}}
\begin{document}
\title{A novel exact cosmological solution of Einstein equations}
\author{David H. Oaknin}\email{d1306av@gmail.com}
\affiliation{}
\begin{abstract}
We present a novel homogeneous and geometrically flat exact solution of Einstein's General Relativity equations for an ideal fluid. The solution, which describes an expanding/contracting hypercylinder, fits well with the observational pillars upon which rely the standard FLRW cosmology and, furthermore, it can naturally solve some of its most outstanding problems.
\end{abstract}
\maketitle

%We present a novel homogeneous, locally isotropic and geometrically flat solution of Einstein's General Relativity %equations for an ideal fluid. We consider a Lorentzian manifold whose metric can be parameterized as $g_{\mu \nu} = %diag\left(1, -a^2(t) r_0^2/r^2, -a^2(t) r_0^2/r^2, -a^2(t) r_0^2/r^2\right)$ in a cartesian comoving grid $(t,x,y,z)$ %in which the fluid is assumed to be at rest $u^{\mu}=(1,0,0,0)$, $t$ denotes cosmological time, $r=\sqrt{x^2+y^2+z^2}$ %is a radial comoving coordinate and $r_0$ is a parametric $length$ scale. This solution, which describes an %expanding/contracting hypercylinder, fits well with the observational pillars upon which relies the standard flat FLRW %cosmology. Furthermore, it naturally solves some of its most outstanding problems.

The Friedmann-Lemaitre-Robertson-Walker(FLRW) exact solution of Einstein's General relativity equations for an ideal fluid describes an expanding/contracting homogeneous and isotropic universe and, hence, it implements the so called Cosmological Principle. This cosmological model, regarded in the literature as the standard cosmology, relies on three solid observational pillars: a) the relation between redshift and distance of nearby astrophysical objects, b) the extremely isotropic microwave blackbody radiation that permeates the whole sky and c) the collected data on primordial densities of light nuclei. The model, whose metric element is given by $ds^2 = dt^2 - a^2(t)\left(\frac{dr^2}{1 - k r^2} + r^2 d\Omega^2\right)$, considers three possibilities: a) $k = 0$ for a spatially flat euclidean universe; b) $k > 0$ for a closed (hyper)spherical universe and c) $k < 0$ for an open hyperbolical universe. The available observational precision data favors the first option, a geometrically flat universe.

Nonetheless, it is widely accepted that the standard FLRW cosmology is an incomplete model. Some unsolved theoretical issues (the flatness problem, the horizon problem,...) and disagreements between its predictions and the observational data (the baryon density needed to fix the data on primordial densities of light nuclei barely accounts for 5\% of the critical mass density needed for a geometrically flat universe) have forced cosmologists to incorporate into the standard FLRW model new speculative ingredients. Namely, a very early stage of cosmological inflation is widely assumed to have preceded the standard evolution of the universe, even though there does not exist yet neither theoretical nor direct observational confirmation of the mechanism that could have driven it. Furthermore, it is necessary to postulate that some mysterious dark energy or cosmological constant of unknown nature accounts for almost 70\% of the energy content of the universe, while another 25\% is due to some kind of cold dark matter. This phenomenological extension of the standard FLRW flat cosmology is known in the literature as concordance $\Lambda$CDM cosmology, for it largely reproduces the precision data collected during the last twenty years \cite{Spergel,Perlmutter}. Yet, some authors have noticed intriguing disagreements \cite{Copi} and also alternative cosmological scenarios has been proposed \cite{Moffat}.

Here we present a novel spatially homogeneous and geometrically flat exact solution of Einstein's General Relativity equations and explore the possibility that it could describe the patch of the universe observable to us. This novel solution, which describes an expanding/contracting hypercylinder, naturally solves the flatness problem of standard FLRW cosmology. In addition, it can fit the observational data upon which relies the standard FLRW cosmology without any need to postulate any kind of mysterious dark energy.

We consider a 4-dim Lorentzian manifold whose metric in a cartesian grid ${\it x} \equiv (x^0,x^1,x^2,x^3)$ we parameterize as

\begin{eqnarray}
\label{the_metric}
g_{0 0}({\it x}) = 1, \hspace{0.2in} g_{0 i}({\it x}) = g_{i 0}({\it x}) = 0, \hspace{0.2in}
g_{i j}({\it x}) = -\frac{r_0^2}{r^2} \left[a^2(t) \Omega_i \Omega_j + b^2(t) \left(\delta_{i j} - \Omega_i \Omega_j  \right)\right],
\end{eqnarray}
where $t \equiv x^0$ is cosmological time, $\{x^i\}_{i=1,2,3}$ are comoving coordinates over the 3-dim spatial sheet, $r \equiv \left(\delta_{i j} x^i x^j\right)^{1/2}$ is a comoving radial coordinate, $\Omega_i \equiv \delta_{i j} \Omega^j = \delta_{i j} \frac{x^j}{r}$ are normalized spatial coordinates and $r_0$ fixes a reference comoving scale. This metric describes and expanding/contracting hypercylinder with comoving radius $r_0$ (see Fig.1). This can be immediately noticed by writing the invariant metric element as

\begin{eqnarray}
\begin{array}{cccc}
ds^2 & = g_{\mu \nu} dx^{\mu} dx^{\nu} = dt^2 - \frac{r_0^2}{r^2} \left[a^2(t) \left(\Omega_i dx^i \right) \left(\Omega_j dx^j \right) + b^2(t) \left(\delta_{i j} - \Omega_i \Omega_j \right) dx^i dx^j \right] \\
%     & = & dt^2 - \frac{r_0^2}{r^2} \left[a^2(t) dr^2 + b^2(t) r^2 \delta_{i j} d\Omega^i d\Omega^j\right] & = \\
\label{the_metric_element}
     & = dt^2 - r_0^2 \left[a^2(t) d{\tilde r}^2 + b^2(t) \left(d\theta^2 + sin^2 \theta  \hspace{0.02in} d\phi^2 \right)\right],
\end{array}
\end{eqnarray}
where ${\tilde r} = ln(r/r_0)$ and $\theta$, $\phi$ are comoving spherical polar coordinates. In order to obtain the last expression we write $dx^i = d(r \hspace{0.025in} \Omega^i) = r \hspace{0.025in} d\Omega^i + \Omega^i \hspace{0.025in} dr$ and then notice the normalization constrain $\delta_{i j} \hspace{0.025in} \Omega^i \hspace{0.025in} \Omega^j = 1$, which implies $\delta_{i j} \hspace{0.025in} \Omega^i \hspace{0.025in} d\Omega^j = 0$. Hence, the dimesionless scale factors $a(t)$ describes the expansion/contraction of the hypercylinder along the radial (axial) direction, while $b(t)$ describes the expansion/contraction of its physical radius.

After some straightforward calculations we can find the components of the Ricci tensor associated to the given manifold,

\begin{eqnarray}
\label{the_Ricci_tensor}
{\it R}_{0 0 }({\it x}) = R_O(t), \hspace{0.15in}
{\it R}_{0 i}({\it x}) = {\it R}_{i 0}({\it x}) = 0, \hspace{0.15in}
{\it R}_{i j}({\it x}) = \frac{r_0^2}{r^2} \left(R_A(t) \Omega_i \Omega_j + R_B(t) \left(\delta_{i j} - \Omega_i \Omega_j\right)\right),
\end{eqnarray}
where $R_O(t) \equiv - \left(\frac{\ddot{a}(t)}{a(t)} + 2 \frac{\ddot{b}(t)}{b(t)}\right)$, $R_A(t) \equiv a(t) \ddot{a}(t) + 2 a(t) \dot{a}(t) \frac{\dot{b}(t)}{b(t)}$ and $R_B(t) \equiv b(t) \ddot{b}(t) + \dot{b}(t)^2 + \frac{\dot{a}(t)}{a(t)} b(t) \dot{b}(t) + r_0^{-2}$ are functions of cosmic time only. The corresponding Ricci scalar ${\it R}  = {\it R}_{\mu \nu} g^{\mu \nu}$ is constant over the spatial sheet,

\begin{equation}
{\it R}({\it x}) =-2 \left(\frac{\ddot{a}(t)}{a(t)} + 2 \frac{\ddot{b}(t)}{b(t)} + 2 \frac{\dot{a}(t)}{a(t)} \frac{\dot{b}(t)}{b(t)} + \frac{\dot{b}(t)^2}{b(t)^2} + \frac{r_0^{-2}}{b(t)^2}\right).
\end{equation}
The components of the Einstein tensor ${\it G}_{\mu \nu}={\it R}_{\mu \nu}-\frac{1}{2} {\it R} g_{\mu \nu}$ are,
therefore, given by:

\begin{eqnarray}
\label{the_Einstein_tensor}
{\it G}_{0 0}({\it x}) = G_O(t), \hspace{0.15in}
{\it G}_{0 i}({\it x}) = {\it G}_{i 0}({\it x}) = 0, \hspace{0.15in}
{\it G}_{i j}({\it x}) = \frac{r_0^2}{r^2} \left(G_A(t) \Omega_i \Omega_j  + G_B(t) \left(\delta_{i j} - \Omega_i \Omega_j\right)\right) ,
\end{eqnarray}
where $G_O(t) \equiv 2 \frac{\dot{a}(t)}{a(t)} \frac{\dot{b}(t)}{b(t)} + \frac{\dot{b}(t)^2}{b(t)^2} + \frac{r_0^{-2}}{b(t)^2}$, $G_A \equiv -\left(2 \frac{\ddot{b}(t)}{b(t)} + \frac{\dot{b}(t)^2}{b(t)^2} + \frac{r_0^{-2}}{b(t)^2} \right) a^2(t)$ and $G_B(t) \equiv -\left(\frac{\ddot{a}(t)}{a(t)} + \frac{\ddot{b}(t)}{b(t)} + \frac{\dot{a}(t)}{a(t)} \frac{\dot{b}(t)}{b(t)}\right) b^2(t)$.

%In General Relativity the Einstein tensor is proportional to the energy-momentum tensor.
For an ideal relativistic fluid the energy-momentum tensor is given by $\label{the_energy_momentum_tensor}
T_{\mu \nu}({\it x}) = \left[\epsilon({\it x}) + P({\it x})\right] u_{\mu}({\it x}) u_{\nu}({\it x}) - P({\it x})
g_{\mu \nu}({\it x})$, where $\epsilon({\it x})$ and $P({\it x})$ are, respectively, the energy density and pressure of the fluid in its proper local frame of reference and $u^{\mu}$ is the velocity four-vector. If we assume that the fluid is at rest in the comoving grid, then $u^{\mu} = \left(1, 0, 0, 0\right)$ and the energy-momentum tensor takes the form,

\begin{equation}
\label{the_energy_momentum_tensor}
{\it T}_{0 0}({\it x}) = \epsilon({\it x}), \hspace{0.15in}
{\it T}_{0 i}({\it x}) = {\it T}_{i 0}({\it x}) = 0, \hspace{0.15in}
{\it T}_{i j}({\it x}) = \frac{r_0^2}{r^2} P({\it x}) \left(a^2(t) \Omega_i \Omega_j + b^2(t) \left(\delta_{i j} - \Omega_i \Omega_j\right)\right) .
\end{equation}
Therefore, Einstein equations ${\it G}_{\mu \nu} + \Lambda g_{\mu \nu} = 8\pi G T_{\mu \nu}$, where for the sake of generality we have included a non zero cosmological constant $\Lambda$, can be exactly solved if both the energy density and the pressure of the fluid are homogeneous over the spatial sheet and evolve in time according to the set of equations

\begin{equation}
\label{the_density}
\epsilon(t) = \frac{1}{8\pi G}\left[2 \frac{\dot{a}(t)}{a(t)} \frac{\dot{b}(t)}{b(t)} + \frac{\dot{b}(t)^2}{b(t)^2} + \frac{r_0^{-2}}{b(t)^2} + \Lambda\right],
\end{equation}
\begin{equation}
\label{the_pressure1}
P(t) = -\frac{1}{8\pi G}\left[2 \frac{\ddot{b}(t)}{b(t)} + \frac{\dot{b}(t)^2}{b(t)^2} + \frac{r_0^{-2}}{b(t)^2} + \Lambda\right],
\end{equation}
\begin{equation}
\label{the_pressure2}
P(t) = -\frac{1}{8\pi G}\left[\frac{\ddot{a}(t)}{a(t)} + \frac{\ddot{b}(t)}{b(t)} + \frac{\dot{a}(t)}{a(t)} \frac{\dot{b}(t)}{b(t)} + \Lambda\right],
\end{equation}
This set of three equations can be solved for $a(t)$, $b(t)$, $\epsilon(t)$ and $P(t)$ once the initial conditions are fixed and the equation of state of the fluid is provided, {\it i.e.} $P(t)=\frac{1}{3}\epsilon(t)$ for a radiation dominated fluid or $P(t)=0$ for a matter dominated fluid.

In order to get analytical insight on the solutions of this set of equations, let us start by considering the limit $r_0^{-1} \ll \left|\dot{b}(t)\right|$ in which the comoving radius of the hypercylinder expands/contracts very fast (indeed, much faster than light). In this limit the term $r_0^{-2}/b(t)^2$ proportional to the curvature of the hypercylinder can be neglected and, hence, we can set $a(t)=b(t)$ and the set of three equations reduces to the usual two equations of the standard FLRW flat cosmology,

\begin{equation}
\label{the_densityFLRW}
\epsilon(t) = \frac{3}{8\pi G}\left[\frac{\dot{a}(t)^2}{a(t)^2} + \frac{\Lambda}{3}\right],
\end{equation}
\begin{equation}
\label{the_pressureFLRW}
P(t) = -\frac{1}{8\pi G}\left[2 \frac{\ddot{a}(t)}{a(t)} + \frac{\dot{a}(t)^2}{a(t)^2} + \Lambda\right],
\end{equation}
In this framework the energy density of the universe evolves roughly as

\begin{equation}
\label{the_densitiy_evolution}
\epsilon(t) = \frac{\epsilon_{0,rad}}{a^4(t)} + \frac{\epsilon_{0,matter}}{a^3(t)},
\end{equation}
(this expression is only qualitatively true, though, as it does not take into account that at high temperatures massive species behave relativistically and contribute to the radiation energy density, while at lower temperatures they can decay into lighter modes) and, therefore, the Hubble expansion rate $H(t) \equiv \dot{a}(t)/a(t)$ evolves as

\begin{equation}
\label{the_Hubble_horizon}
H^2(t) = \frac{8\pi G}{3} \left(\frac{\epsilon_{0,rad}}{a^4(t)} + \frac{\epsilon_{0,matter}}{a^3(t)}\right)
- \frac{\Lambda}{3}.
\end{equation}
Hence, the radiation density controls the expansion rate of the universe at very early times and the matter density takes over later on. If the cosmological constant is not zero, $\Lambda \neq 0$, it eventually becomes dominant at some point. In the concordance $\Lambda$CDM model the cosmological constant becomes dominant only at present time.

Let us now come back to the full set of equations (\ref{the_density}, \ref{the_pressure1}, \ref{the_pressure2}) that describe the dynamics of the hypercylindrical cosmological framework. For the sake of simplicity we rewrite the equations in terms of $\tau \equiv H_0 t$, where $H_0$ is the current best observational estimate (within the framework of the standard FLRW cosmology) of the Hubble expansion rate at present. As we are interested in exploring the possibility that the novel term proportional to the curvature of the hypercylinder $r_0^{-2}/b(t)^2$ could mimic the contribution of a non-zero cosmological constant, from now on we fix $\Lambda = 0$ in these equations:

\begin{equation}
\label{the_densityB}
\epsilon(\tau) = \frac{H_0^2}{8\pi G}\left[2 \left(\frac{1}{a(\tau)} \frac{da(\tau)}{d\tau}\right) \left(\frac{1}{b(\tau)} \frac{db(\tau)}{d\tau}\right) + \left(\frac{1}{b(\tau)} \frac{db(\tau)}{d\tau}\right)^2 + \frac{\left(H_0^{-1}/r_0\right)^2}{b(\tau)^2}\right],
\end{equation}
\begin{equation}
\label{the_pressure1B}
P(\tau) = -\frac{H_0^2}{8\pi G}\left[2 \frac{1}{b(\tau)} \frac{d^2b(\tau)}{d\tau^2} + \left(\frac{1}{b(\tau)} \frac{db(\tau)}{d\tau}\right)^2 + \frac{\left(H_0^{-1}/r_0\right)^2}{b(\tau)^2}\right],
\end{equation}
\begin{equation}
\label{the_pressure2B}
P(\tau) = -\frac{H_0^2}{8\pi G}\left[\frac{1}{a(\tau)} \frac{d^2a(\tau)}{d\tau^2} + \frac{1}{b(\tau)} \frac{d^2b(\tau)}{d\tau^2} + \left(\frac{1}{a(\tau)} \frac{da(\tau)}{d\tau}\right) \left(\frac{1}{b(\tau)} \frac{db(\tau)}{d\tau}\right)\right],
\end{equation}
For $r_0 \simgt H_0^{-1}$ the curvature term is negligible at very early times and, therefore, the hypercylindrical cosmological framework  is indistinguishable from the standard FLRW flat cosmology (\ref{the_densityFLRW}), (\ref{the_pressureFLRW}) during the early stages of the history of the universe. Only at present time the curvature term becomes non-negligible and the hypercylindrical cosmological model diverges from the standard FLRW flat cosmology. In particular, the two scale factors do not remain equal anymore $a(t) \neq b(t)$ and, therefore, neither the Hubble expansion rates $H_a(t) \equiv \dot{a}(t)/a(t)$ and $H_b(t) \equiv \dot{b}(t)/b(t)$ remain equal.

The late evolution of the scale factors within this novel cosmological framework is plotted in Fig. 2 ($a(t)$ - red line; $b(t)$ - blue line) as a function of cosmological time for $H_0^{-1}/r_0 = 0.53$ and initial conditions that correspond to those of the standard FLRW cosmology at the instant of decoupling. The solution is compared to the evolution of the scale factor in the concordance $\Lambda$CDM model (black line) and in the matter dominated flat FLRW cosmology (grey line) for the same given initial conditions. These initial conditions were normalized such that within the framework of the matter dominated flat FLRW cosmology the scale factor at time $t = H_0^{-1}$ would be equal to 1. For the sake of simplicity, in all the three theoretical frameworks the contribution of radiation was neglected since the instant of decoupling (that is, we set $P(t)=0$). It can be seen that the evolution of the scale factor along the axial direction in the hypercylindrical model fits very well with the phenomenological $\Lambda$CDM model, without any need to postulate any kind of dark energy. The value $H_0^{-1}/r_0 = 0.53$ used here is only an arbitrary choice and definitely not the output of any optimization fit. Such an optimization will be carried separately. Let us remark that $b(t)$ is the scale factor upon a 2D sphere that contains the {\it plane} of our Galaxy, while $a(t)$ describes the dynamics along the orthogonal spatial direction. In Fig. 3 is plotted the evolution of the energy density in this  hypercylindrical model in units of $\epsilon_{cr} \equiv \frac{3 H_0^2}{8\pi G}$.

The so-called {\it flatness problem} of standard FLRW cosmology gets immediately solved within the hypercylindrical cosmological framework that we have just described, for its spatial manifold is geometrically flat for any value of the cosmological energy density $\epsilon_0$ or the radius $r_0$ \footnote{The geometrical {\it flatness} of the hypercylindrical universe can be intuitively understood by noticing that a flat piece of paper can be rolled into a cylinder without modifying the distances between points.}. Hence, the model can naturally fit the location of the acoustic peaks in the spectrum of primodial cosmic microwave temperature anisotropies over the sky. Moreover, the hypercylindrical universe can also accommodate recent claims of a preferred direction in the sky map of primordial temperatures anisotropies in the cosmic microwave background \cite{Copi}.

In Fig. 4 we plot the evolution of the same hypercylindrical model of Fig. 2 and Fig. 3 well beyond the present time.
The scale factor $b(t)$ collapses in finite time and drives the exponential expansion of the scale factor along the axial direction $a(t)$. This behavior suggests that the collapse of large additional extra dimensions in the very early stages of the universe might have driven a kind of inflation along the remaining present three spatial dimensions. This issue will be explored separately. 

The hypercylindrical solution can also be smoothly matched to, for example, a closed FLRW universe (hypersphere) with comoving radius $r_0$. We can then speculate about about the universe as a tree (see Fig. 5) of hierarchical structures.

%The novel exact solution of Einstein's General Relativity equations for a relativistic ideal fluid that we have just %presented is a generalization of the exact solution of the special relativistic hydrodynamics equations presented in %\cite{Oaknin}. The new solution is an answer the following thought experiment. Consider a linear array of emitters at %rest in a grid of inertial observers and an additional observer that can move along this line. From the point of view %of this last observer, all emitters (even those beyond its causal horizon) respond immediately when it accelerates. It %cannot know immediately about this response, but if it waits enough it can know that even the furthest emitters %responded to its changes. Obviously, there is not any violation of causality because our observer is not inertial. On %the other hand, within the framework of General Relativity all frames of reference - inertail or not -  are %equivalent.
%That is, General Relativity seems to allow immediate responses beyond the causal horizon. In a linear 1D array it is %quite obvious to characterize these motions. The solution presented in this paper characterizes the same kind of %responses in a spherically symmetric set up.

%{\bf Acknowledgments}
%I am gratefully indebted to my teacher Ana Oaknin~(z.l) for her strong and continuous
%support and for her example of honesty and dedication.

%\includegraphics[scale=0.5]{cylinder.eps}
\begin{figure}[htbp]
\begin{center}
\scalebox{0.5}{\psfig{file=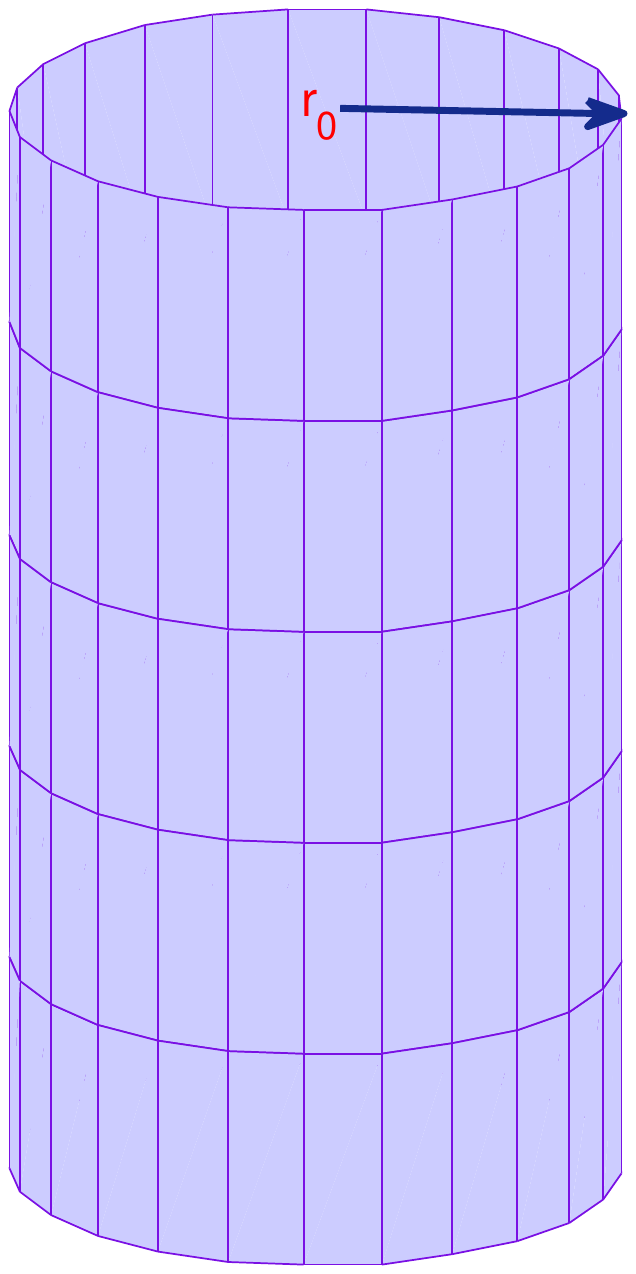}}
\end{center}
\caption{Comoving spatial sheet associated to metric (\ref{the_metric}). Each horizontal circle
represents indeed a 3D sphere.}
\label{figure: Fig.1}
\end{figure}

\begin{figure}[htbp]
\begin{center}
\scalebox{0.5}{\psfig{file=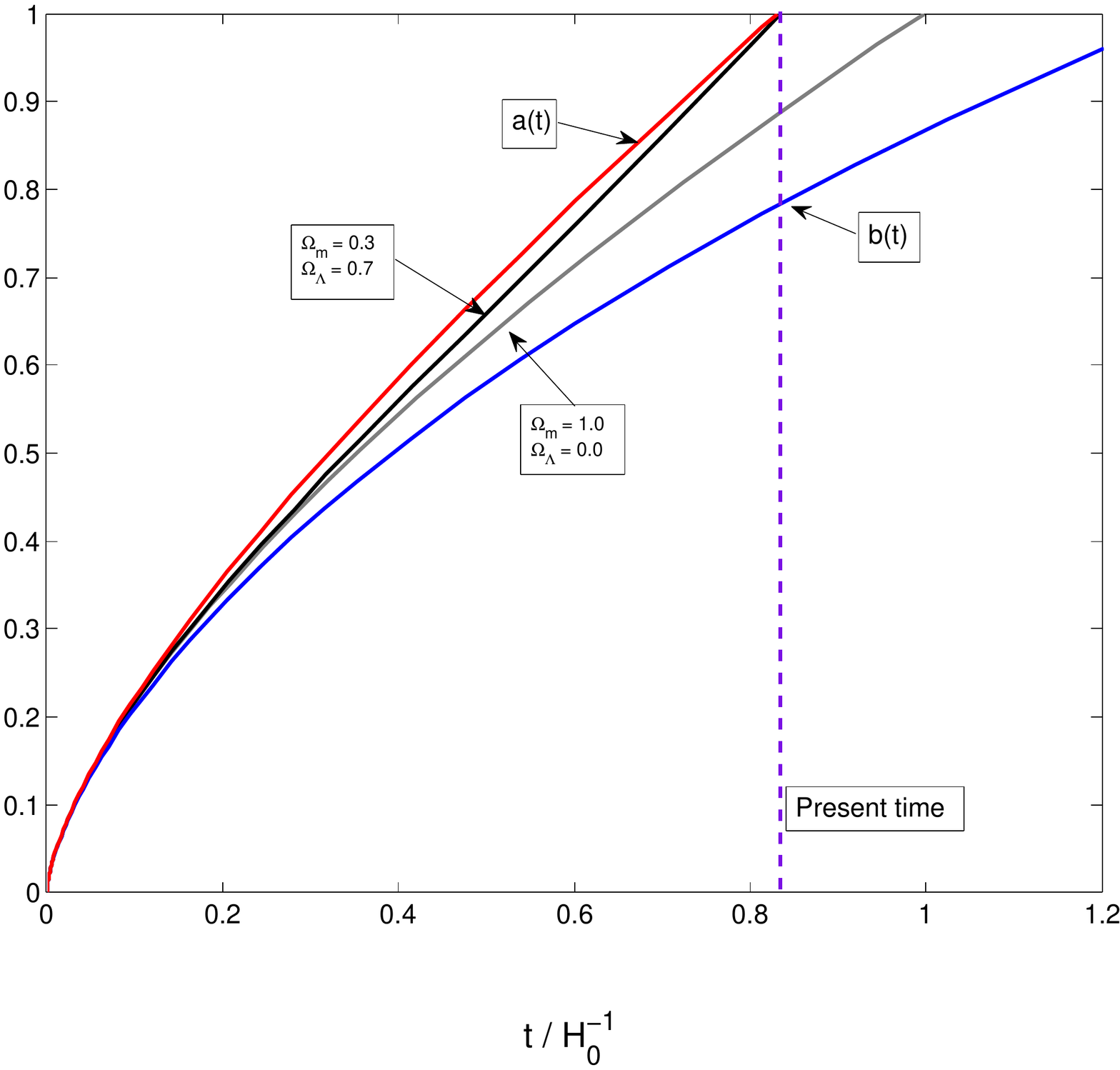}}
\end{center}
\caption{Evolution of the scale factor within three different cosmological models for identical initial conditions at the instant of decoupling: a) matter dominated flat FLRW universe, $\Omega_{m} = 1$ (gray line); b) $\Lambda$CDM with $\Omega_{m} = 0.3$ and $\Omega_{\Lambda} = 0.7$ (black line); c) hypercylindrical model with $H_0^{-1}/r_0 = 0.53$, red line -- evolution of the scale factor along the axial direction, blue line -- evolution of the scale factor along the orthogonal 2D manifold. The purple line signals the present age of the universe. The hypercylindrical model fits well with the phenomenological $\Lambda$CDM model without any need to postulate any kind of dark energy.}
\label{figure: Fig.2}
\end{figure}

\begin{figure}[htbp]
\begin{center}
\scalebox{0.5}{\psfig{file=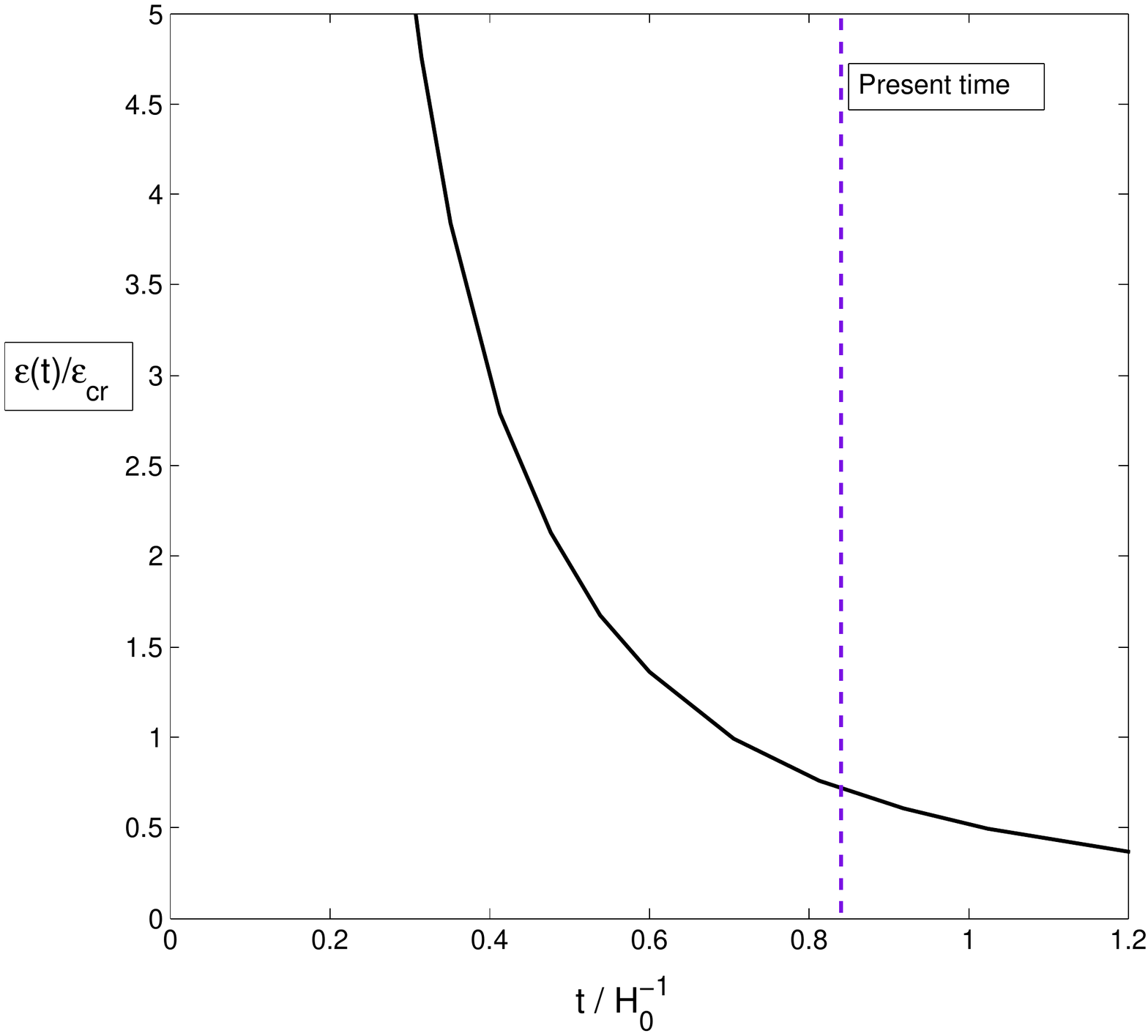}}
\end{center}
\caption{Evolution of the energy density in the hypercylindrical model. The purple line signals the present age of the universe.}
\label{figure: Fig.3}
\end{figure}

%\begin{figure}[htbp]
%\begin{center}
%\scalebox{0.5}{\psfig{file=cylinder_ball.pdf}}
%\end{center}
%\caption{The comoving (hyper)cylinder can be smoothly matched to a comoving hypersphere in order to avoid the %singularity at the origin. The red line represents the comoving surface of last scattering for an observer located %close to the origin of the comoving grid.}
%\label{figure: Fig.4}
%\end{figure}

\begin{figure}[htbp]
\begin{center}
\scalebox{0.5}{\psfig{file=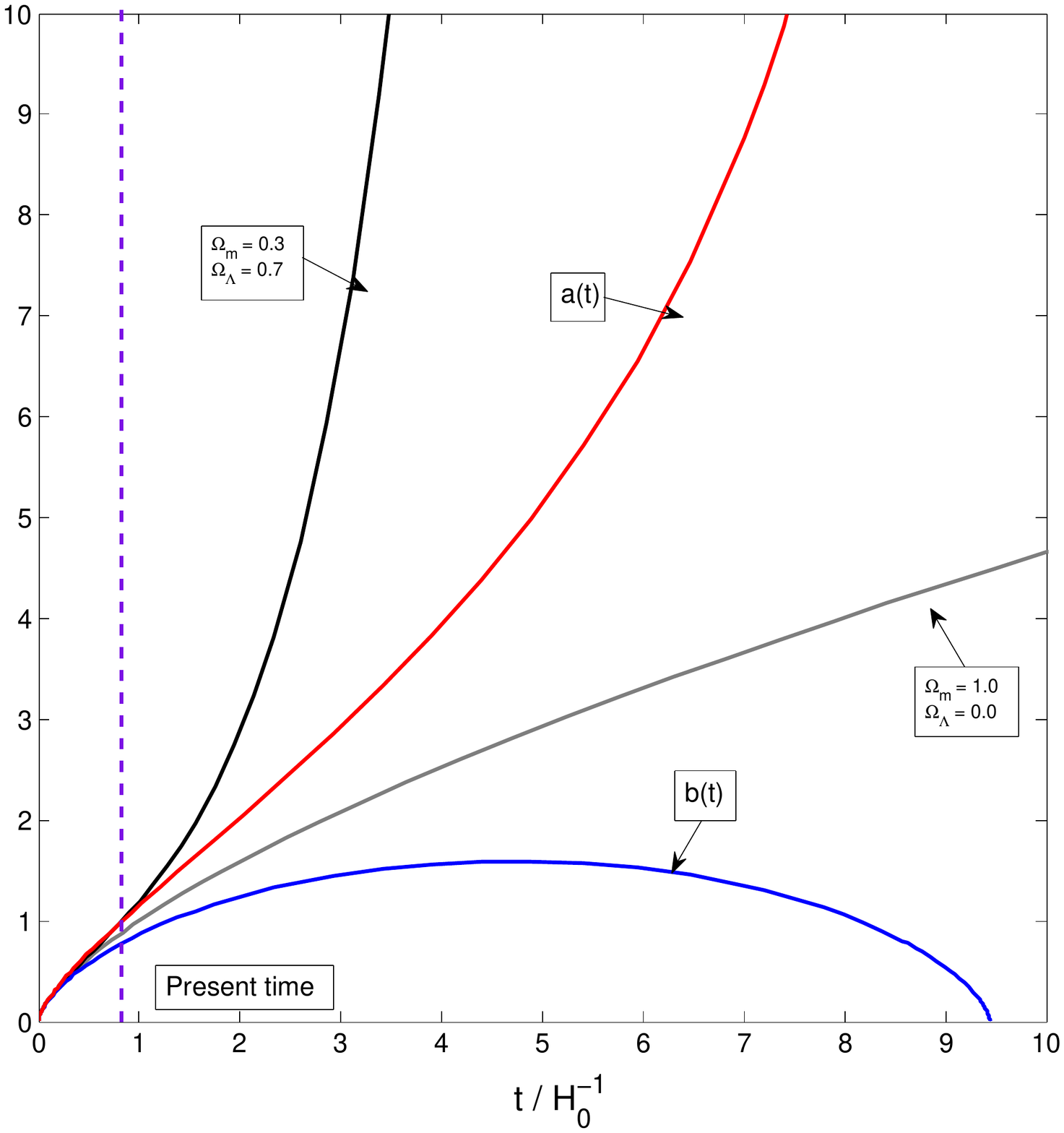}}
\end{center}
\caption{Evolution of the scale factor well beyond the present time within three different cosmological models with identical initial conditions at the instant of decoupling: a) matter dominated flat FLRW universe, $\Omega_{m} = 1$ (gray line); b) $\Lambda$CDM with $\Omega_{m} = 0.3$ and $\Omega_{\Lambda} = 0.7$ (black line); c) hypercylindrical model with $H_0^{-1}/r_0 = 0.53$, red line -- evolution of the scale factor along the axial direction, blue line -- evolution of the scale factor along the orthogonal 2D manifold. The purple line signals the present age of the universe.}
\label{figure: Fig.2}
\end{figure}

\begin{figure}[htbp]
\begin{center}
\scalebox{0.5}{\psfig{file=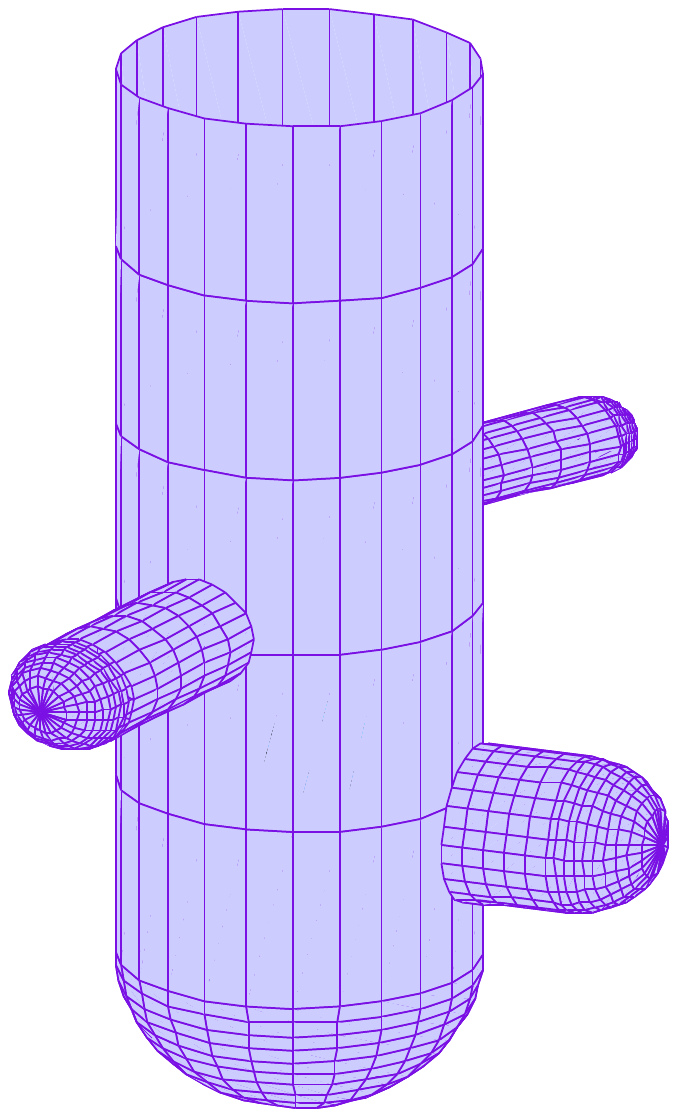}}
\end{center}
\caption{Hypothetical structure of the universe. Our patch of the universe might be one of the hypercylinders that asymptotically match into a larger similar structure. Smaller hypercylinders could also evolve from our visible patch.}
\label{figure: Fig.5}
\end{figure}

\end{document}